# Preparing fMRI Data for Statistical Analysis

Alfonso Nieto-Castanon [1][2]

[1] Department of Speech, Language, and Hearing Sciences, Boston University, Boston, Massachusetts, United States of America

[2] Department of Brain and Cognitive Sciences, Massachusetts Institute of Technology, Cambridge, Massachusetts, United States of America

**Abstract**

This chapter describes several procedures used to prepare fMRI data for statistical analyses. It includes the description of common preprocessing steps, such as spatial realignment, coregistration, and spatial normalization, aimed at the spatial alignment of all fMRI data within- and between- subjects, as well as several denoising procedures aimed at minimizing the impact of common noise sources, including physiological and residual subject motion effects, on the BOLD signal time series. The chapter ends with a description of quality control procedures recommended for detecting potential problems in the fMRI data and evaluating its suitability for subsequent statistical analyses.

## 1. Introduction

Blood oxygenation level dependent (BOLD) signal fluctuations reflecting underlying neuronal activation, commonly the focus of statistical analyses of fMRI data, are shadowed by an abundance of other additional, often artifactual, sources of BOLD signal variability. Without careful control of these additional sources, statistical analyses of fMRI data would be drowned in noise, leading to analyses with limited power and low replicability.

Many of these additional factors affecting the BOLD signal are well known. They include subject motion in the scanner, spatial misalignment caused by scanner imprecisions or differences in subjects' anatomy, temporal fluctuations related to cardiac and respiratory cycles, acquisition time differences between individual slices, and thermal and drift noise from the scanner, among others, and they can have a considerable impact on both spatial and temporal properties of the fMRI data.

This chapter describes the steps typically involved in preparing fMRI data for statistical analyses, focusing on removing or minimizing these well-known sources of variability from the BOLD signal and identifying potential problems in the data before proceeding to statistical analyses. The first section (*preprocessing*) describes initial processing steps focusing on the spatial properties of both the functional and anatomical images. Some of these steps act, for example, to identify and correct misalignment across these images caused by subject motion within the scanner, magnetic field inhomogeneities, or by anatomical differences between subjects. The second section (*denoising*) focuses on the temporal domain of the BOLD signal. It describes additional processing steps acting to identify and remove remaining sources of temporal variability in the BOLD signal, such as cardiac, respiratory and other physiological factors, or residual effects caused by subject motion. The last section (*quality control*) describes common procedures used to detect potential problems in the fMRI data or in any of the preprocessing and denoising steps, and ultimately evaluate the suitability of the resulting data for subsequent statistical analyses.

## 2. Preprocessing

Each functional run in the scanner consists of a series of 3D images or scans acquired sequentially, typically with a fixed sample rate in the order of 1 Hz, and lasting for a few minutes. Each image is organized as a three-dimensional volume encoding the BOLD signal sampled at a uniform array of individual voxels (three-dimensional pixels) covering a typically-fixed region of space inside the scanner. The dimensionality of the fMRI data from a single functional run normally consists of several hundred thousand voxels each scanned at several hundred timepoints.

Initial preprocessing steps focus on *intrasubject* coregistration, compensating for a subject's head motion across different scans, and *intersubject* coregistration, the alignment of all of the subjects' brains in a way that preserves their main common anatomical features. Correction of other scanner-related artifacts, such as those produced by inhomogeneities in the magnetic field, or differences in the time of acquisition of individual slices, are often also addressed during these initial preprocessing steps, either jointly or separately from the above procedures.

### 2.1 Susceptibility Distortion Correction (SDC)

Differences in magnetic susceptibility across diverse tissue classes alter the precise homogeneity of the MR scanner field once a subject is introduced in the scanner. These inhomogeneities are stronger in areas close to tissue-air boundaries and produce both spatial distortions as well as signal dropout in the fMRI data. These effects are readily evident in many EPI images, for example, in inferior areas of the frontal lobe close to the nasal sinus. Spatial distortions arising from field inhomogeneities are particularly problematic along the phase-encoding direction (often the anterio-posterior direction, or AP) in an EPI sequence, and they have no significant effects on other spatial directions in an EPI sequence or on other acquisition sequences (e.g. T1-weighted used for anatomical scans).

While there has been some work toward methods that allow a purely data-driven correction of these spatial distortions [1], most current approaches rely on the acquisition of a separate scan attempting to quantify the precise distortions in the magnetic field within the scanner [2-3]. This separate *fieldmap* acquisition is performed prior to the fMRI runs. One approach relies on the acquisition of two sequential images acquired with different echo times, where by analyzing the phase difference between the two images a voxel-displacement map characterizing the magnitude and direction of the expected spatial distortions can be constructed. Another approach relies on the acquisition of two images acquired with opposite phase encoding directions (e.g. AP/PA), where the expected distortions can be quantified by comparing the opposite spatial deformations present in these two images (see Figure 1).

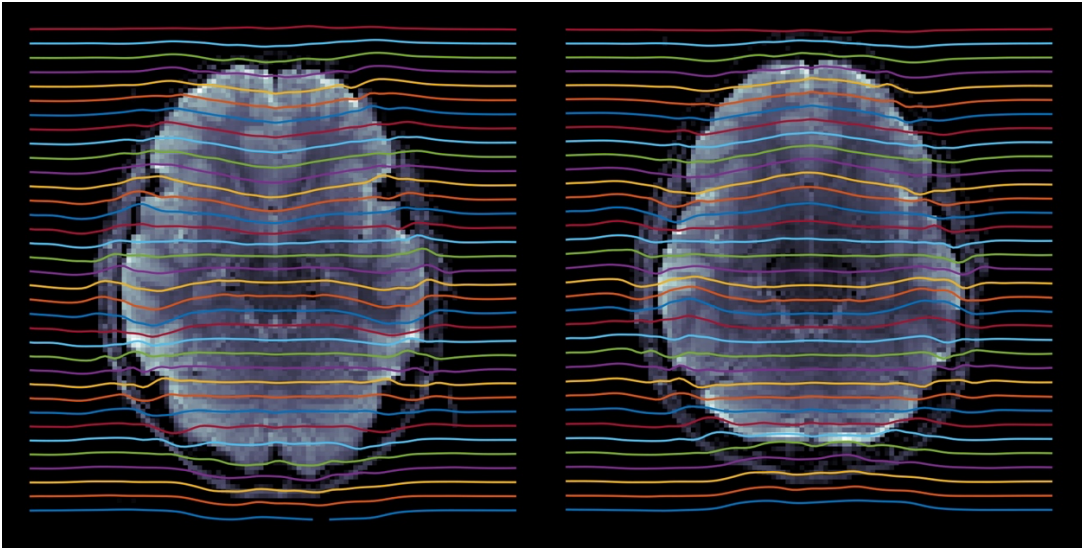

**Fig. 1** Example of functional data (background image) and estimated voxel-displacement map (color lines), for an axial slice acquired with two opposite phase encoding directions (AP left, PA right)

Once a voxel-displacement map is estimated, functional images can be spatially warped to recover the undistorted geometry. For a more efficient correction this warping can be performed simultaneously during realignment (see motion correction section below) to better account for expected changes in the magnitude of these distortions as the subject's head moves during the scanning session.

## 2.2 Motion Correction

Subject motion in the scanner produces large fluctuations in the BOLD signal. Part of those fluctuations are simply due to the average BOLD signal intensity being quite different across different areas and across different tissue classes, so even small movements translate to relatively large changes as the signal time series at any given voxel is being sampled from different locations in the brain.

Considering that the brain is relatively rigid, a first-order correction of motion-induced changes in fMRI data can be achieved in two steps, in a process often referred to as *realignment* or *intramodality coregistration*. In the first step, the functional data are analyzed to estimate how exactly a subject moved in the scanner, for example by comparing the images acquired at each individual timepoint $I_n$ to the first image or other common reference image $I_0$ [4]. A rigid-body transformation matrix $M_n$ can be used to uniquely characterize the motion between these two images. This transformation matrix is encoded as a 4x4 matrix that maps 3D voxel coordinates $x_0$ in the reference image to the 3D voxel coordinates $x_n$ in the n-th image:

$$\begin{bmatrix} x_n \\ 1 \end{bmatrix} = M_n \cdot \begin{bmatrix} x_0 \\ 1 \end{bmatrix}$$

in a way that minimizes some measure of dissimilarity between the two images, e.g.:

$$\min_{M_n} \sum_{x_0} |I_n(x_n) - I_0(x_0)|^2$$

For simplicity, rigid-body transformation matrices $M_n$ characterizing subject motion across time are usually represented by a series of six parameters, three representing translations and three representing rotations along each axis. Unfortunately, these reduced representations are not uniquely defined. Different software packages use different conventions for their definition. For example, SPM [5] and CONN [6] use rotations around the point with (0,0,0) world-space coordinates, where rotations are represented in radians and applied sequentially in the order Z-Y-X, while FSL [7], AFNI [8], or fMRIPrep [9] use each different definitions of the center of rotation, units, and order of application of the different rotations. This disparity results in different sets of reduced motion parameters for the same data, as well as motion parameters that depend on the data's original orientation, so reduced representations of rotation and translation parameters, while widely used to characterize subject motion, should always be interpreted with care.

Once the subject motion in the scanner has been estimated, the second step is simply to apply an opposite rigid-body transformation to the fMRI images at each timepoint to counteract those movements, effectively coregistering all scans of a subject to a common reference. Most modern imaging file formats, such as NIFTI or ANALYZE, allow separate encoding of the raw imaging data from the information regarding the orientation of these data. The orientation of an image is encoded by an affine *voxel-to-world* matrix $T_n$ (see Figure 2), characterizing the transformation from 3D indices $i$ of a voxel in this image to the position $x$ of that same voxel in the world in a conventional reference frame (e.g. with x/y/z coordinates in mm units and encoding left-to-right, anterior-to-posterior, and inferior-to-superior directions of the head):

$$\begin{bmatrix} x_n \\ 1 \end{bmatrix} = T_n \cdot \begin{bmatrix} i_n \\ 1 \end{bmatrix}$$

In these cases, once the subject motion has been estimated, coregistration of the images can be accomplished without the need to interpolate the functional data simply by modifying the voxel-to-world matrices $T_n$ to include the necessary subject-motion correction:

$$T_n^* = M_n^{-1} \cdot T_n$$

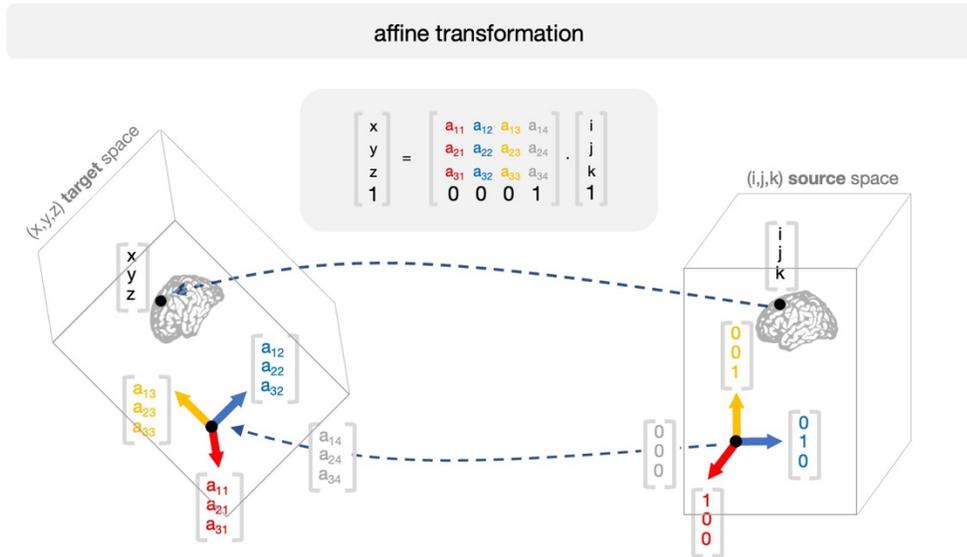

**Fig. 2** Schematic of affine map between voxel indices in a 3D volume (right) and 3D world coordinates (left), characterizing the position and orientation of the brain in a standard reference frame.

Subject motion also interacts with susceptibility-induced distortions. In particular, the distribution of inhomogeneities in the scanner field created by the presence of the subject changes when the subject moves, altering the form of the resulting spatial distortions in functional images. To address this interaction, a first-order linear approximation to the changes in the magnetic field associated with subject motion is estimated from the functional images, which is then used to update the voxel-displacement map estimate and jointly correct susceptibility distortions and subject motion individually at each timepoint [10].

Despite these motion correction procedures a large number of residual effects of motion still remain in the BOLD time series. Among the largest factors that are not addressed by the motion-correction procedures described above are:

- Slice timing interactions: subject motion occurs at any point during the acquisition, so it can affect different slices of the functional data differently, depending on when exactly those slices were acquired (see Slice Timing Correction section below). This causes images to change with motion in a way that cannot be simply represented by a rigid-body transformation. A more efficient correction would take into account the interaction between subject motion and other factors such as intrascan differences in slice acquisition times [11-12]
- Spin history effects: as a subject moves in the scanner, the exact location of the tissue that is excited by an individual slice RF pulse varies. The state of relaxation of the tissue when a new pulse arrives depends on the time between successive pulses, so if the tissue is not exactly aligned across successive acquisitions the magnetization state of the tissue will covary with motion, resulting in "spin history" effects that can last several scans [13-14].
- Nonrigid-body noise: motion correction can introduce motion-correlated BOLD signal artifacts when applied to noise sources that do not accompany head movements (e.g. Nyquist ghost artifacts)
- Algorithm inaccuracies: inaccuracies in the estimation of subject motion parameters, as well as edge effects and inaccuracies when resampling the functional data [15], can result not only in a failure to fully remove the modeled motion-related artifacts, but they can also potentially introduce motion-correlated noise in the BOLD signal.

## 2.3 Slice Timing Correction (STC)

In a typical continuous functional acquisition, the BOLD signals from tissue within a single planar 2D slice are measured almost simultaneously using a fast sequence of RF pulses, and the measured electromagnetic signals at the scanner receiver coils are combined to create a single 2D image. These individual images, for example axial slices, are then stacked to form a 3D image, and the process is repeated at a fixed rate (e.g. repetition time TR = 2 s) to generate the complete fMRI data. This procedure results in slightly different acquisition times of the BOLD signal across different slices. The relative importance of correcting for interslice acquisition time differences depends on the type of planned statistical analysis of the fMRI data. For example, in an analysis of low-frequency fluctuations in the BOLD signal, interslice acquisition time differences may not be of great concern. On the other hand, in a fast event related design, where knowing the precise timing of each individual event is necessary, an appropriate STC may be crucial.

Interslice differences in acquisition times can be corrected a posteriori by temporally shifting the BOLD time series at each voxel an amount counteracting the delay between the acquisition time of this voxel vs. that of a predefined reference slice. For example, Fourier-based algorithms resample the BOLD signal at the desired times using sinc-interpolation [16] (see Figure 3), where a fractional delay $\tau$ in a time series *f[n]* would be corrected using a transformation of the form:

$$f^*[n] = f[n - \tau] \approx \frac{1}{N} \sum_{w=1}^{N} \sum_{m=1}^{N} f[m]\, e^{-j2\pi\,(w-[N/2])(m-n+\tau)/N}$$

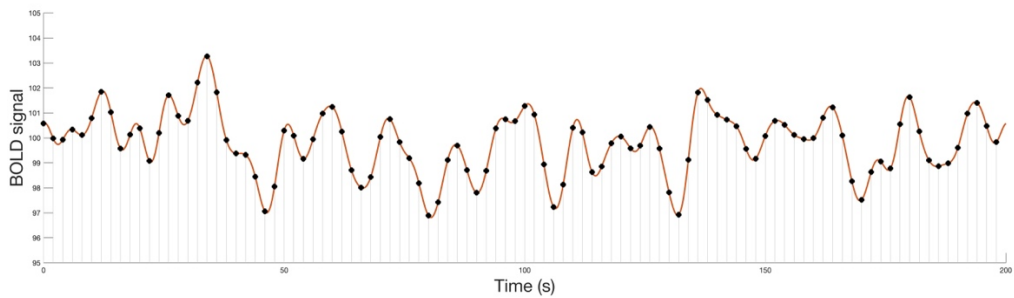

**Fig. 3** Example of sinc-interpolation (continuous red line) of BOLD discrete time series (black dots)

Temporal interpolation may be limited in cases where the TR is long or the acquisitions are sparsely sampled. It can also adversely extend the influence of individual outlier events on the BOLD signal. Rather than correcting the BOLD data directly, other approaches choose instead to model these delays directly during the statistical analysis, either explicitly taking into account the acquisition delay at each voxel or by including a general term (e.g. hemodynamic response function derivatives) that accommodates uncertainty in the precise timing of the events modeled [17].

The procedures used for motion correction and slice timing correction have conflicting assumptions. Motion correction estimation and resampling procedures consider motion at the level of the entire volume, treating motion as a rigid body transformation between volumes and disregarding timing differences between slices, while slice timing correction interpolation procedures assume that a voxel remains stationary, locked to the same location in the brain and disregarding potential motion between the scans. Since most imaging packages, including SPM, CONN, FSL, AFNI, and fMRIPrep, deal with these two sources separately, there is some debate regarding the optimal order of these two preprocessing steps [18]. In general, the recommendation is to first apply the procedure that is suspected to have resulted in larger distortions in the functional data, depending on a dataset sample or acquisition details. For example motion correction may be applied first in a study with large observed levels of subject motion, while STC may be applied first in a study with long TR.

### 2.4 Coregistration

Depending on the planned statistical analyses of fMRI data it may be convenient or necessary to combine information from different modalities. For example, a researcher may define regions of interest on the structural T1-weigthed scan of a subject, where anatomical features are more clearly observable, and then wish to perform statistical analyses of EPI functional data at those same locations. As the subject may have moved between the two scans the same procedures used for motion correction are used to coregister the data from a subject across different modalities.

The main difference between realignment, or intramodality coregistration, and (intermodality) coregistration is that the latter procedures require more general measures of similarity/dissimilarity between the two images, as the relationship between the intensity values at the same anatomical location in two different modalities will naturally be more complex than the relationship between two images of the same modality (see Figure 4). Information theoretical measures such as mutual information and entropy correlation can be used to measure the similarity between two images when the intensity profiles of the different tissue classes in these images may vary largely across the two different modalities or MR contrasts [19-22].

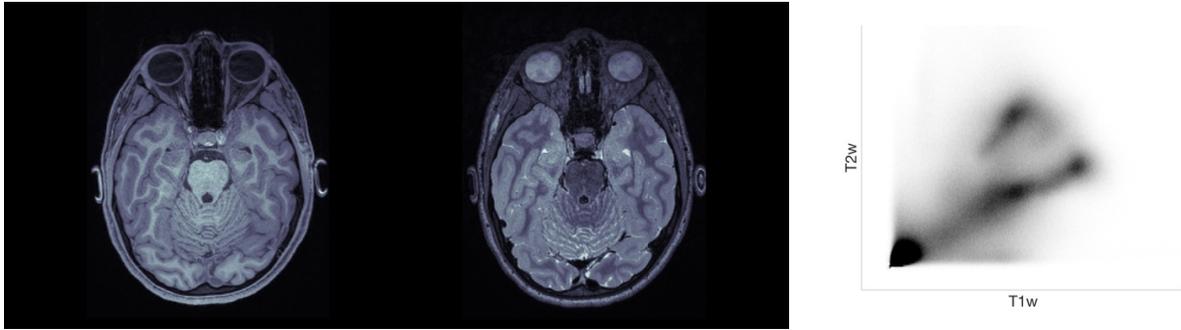

**Fig. 4** Example of an axial slice MR acquired with two different contrasts (images T1-weighted left, T2-weighted right), and the complex association between intensity values in the two images (right scatterplot)

Some of the main difficulties or limitations in the effectiveness of intermodality coregistration arise from potentially large differences in noise or artifactual sources across the two modalities. For example, intensity nonuniformity artifacts in the images can negatively affect the quality of coregistration [23]. In addition, spatial distortions caused by inhomogeneities in the scanner field and discussed in the SDC section above, are specific to EPI images, not affecting other imaging modalities such as T1-weighted normally used to acquire anatomical scans. If these spatial distortions have not been properly corrected in the fMRI data, due for example to a lack of fieldmap scans, coregistration of functional and anatomical data will produce subpar results as the two images cannot be properly aligned using simple rigid-body or affine transformations.

**2.5 Spatial normalization**

Most statistical analyses of fMRI data require combining information from multiple subjects. To that end it is common to assume that there is some form of identity or one-to-one mapping between locations in the brain of different subjects in a way that is invariant to anatomical differences and different shapes and sizes of their brains and heads.

In subject-specific region of interest (ss-ROI) analyses, this mapping is implicitly characterized by one or several common areas that are delineated in the brain of each subject, using either anatomical landmarks or functional localizer tasks [24]. These ROIs then act as the basic units of statistical analyses, where the BOLD signal is first aggregated across all voxels within an ROI, and then combined across subjects. In this way, regions of interest provide a common reference with relatively lax assumptions about the precise spatial localization of function across diverse subjects.

In contrast, in voxel-level analyses individual voxels are the basic units of statistical analyses. Intersubject coregistration of individual voxels relies on spatial normalization, a procedure that warps each subject's brain to match a common template or reference space, although the details of spatial normalization vary depending on whether voxel-level analyses are performed on the cortical surface or across the entire volume.

In surface-based analyses, where the statistical analyses focus only on voxels within the cortical surface of each subject, the mapping between locations in the brain of different subjects is explicitly characterized by matching anatomical features of each subject's cortical surface to a common template [25-26]. The procedure is performed in two steps. First, points in the cortical surface of a subject are mapped to points on a spherical surface in a way that minimizes metric distortions, so that the distance between two points on the cortical surface is similar to the geodesic distance between the locations on the spherical surface where these points are mapped. A map representing this subject's anatomical features (cortical curvature values, with negative values for points within cortical sulci and positive values for points within cortical gyra) is then projected to the spherical surface and warped to match a common template of these features in a standard space (e.g. fsaverage space). These two transformations combined provide a one-to-one mapping between points in each subject's cortical surface and points on a reference surface coordinate system.

For voxel-level analyses encompassing the entire brain, the most common form of intersubject coregistration relies on warping brain images of each subject to match a common template in a standard reference space, such as MNI (Montreal Neurological Institute). In practice, a relatively large number of different matching algorithms and different templates are similarly considered to provide a transformation to a common MNI reference space, even if consistent differences remain between different procedures [27-29]. Template or reference images can be defined from an average of T1-weighted images similarly normalized, such as the ICBM template [30], or from similar average images for other modalities or MR contrasts [31]. More detailed templates, independent of image modality, can also be defined from a set of spatial probability maps of different tissue classes, such as gray matter, white matter, cerebrospinal fluid, soft-tissue, bone, and air [32] (see Figure 5).

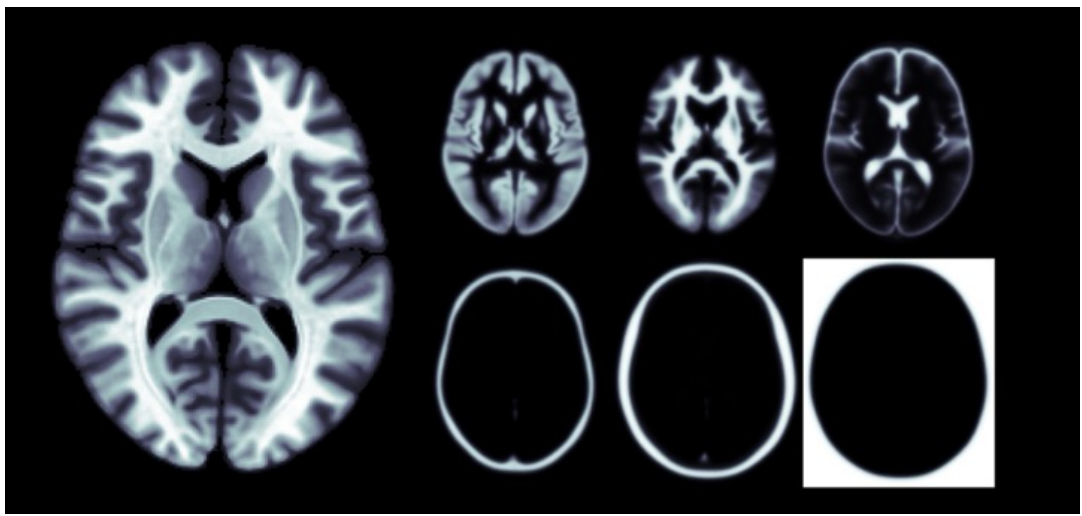

**Fig. 5** Example of an axial slice T1-weighted template (left; ICBM 2009c nonlin asym), and a tissue probability map template (right; SPM IXI-549 TPM)

Modern algorithms can simultaneously segment and normalize images from different modalities with the aid of tissue probability maps that represent the a priori likelihood of different tissue classes over each individual voxel in an image [32-33]. Warping algorithms have also evolved over time, with initial deformation models relying on a small set of basis functions, such as polynomials, discrete cosine, b-spline, or radial basis functions [34-39], while current deformation models [27-28,40] typically rely on some variation of the general framework of diffeomorphic transformations [41-42]. Under this framework deformations are defined by the continuous integration of smooth velocity fields, and the allowable extent of the deformation is effectively controlled by regularization terms during the optimization procedure. This ensures that the resulting transformations preserve the image topology [43], maintaining a one-to-one mapping between points in each subject's brain and points in the target reference space.

Regarding the practical application of spatial normalization in the context of the analysis of fMRI data, there are two different strategies that are used to jointly normalize the anatomical and functional data, which are referred to as *direct* and *indirect* normalization, respectively. In a *direct* normalization procedure, the functional and anatomical data are each normalized separately. In contrast, in an *indirect* normalization procedure the functional and anatomical images are first coregistered, and then only the anatomical image is normalized, while the functional images are instead directly transformed using the same spatial transformation estimated from the anatomical data. While indirect normalization is faster, and it can potentially take advantage of the reduced noise, higher spatial resolution, or higher tissue contrast available in T1-weighted anatomical scans to produce a more accurate functional normalization, in practice the quality of the results relies on the accuracy of the coregistration between the functional and anatomical images, which can be negatively affected by EPI susceptibility distortions (see the coregistration section above). The general recommendation is to use indirect normalization when high-quality fieldmaps are available and EPI susceptibility distortions can be accurately corrected, and use direct normalizations when fieldmaps are not available or when there are significant residual mismatches between functional and anatomical images, as the direct normalization of the functional data can serve to minimize some of those distortions [44].

## 2.6 Spatial smoothing

The last step when preprocessing fMRI data is to spatially smooth the functional images. Smoothing is implemented by a spatial convolution operation with a three-dimensional isotropic Gaussian kernel, and the amount of smoothing is controlled by the width of this kernel, often specified by its full width half maximum (FWHM) value (e.g. 8 mm).

Smoothing serves multiple purposes. First, it can be expected to increase the BOLD signal-to-noise ratio (SNR). A classical matched filter approach determines that an optimal spatial filter maximizing the BOLD signal SNR would have a spatial frequency response equal to the ratio between the energy of the expected BOLD response and that of the expected noise covariance. From that perspective, spatial smoothing is simply a matched filter optimized to detect BOLD responses that are relatively smooth, with the kernel FWHM value controlling the expected extent of these responses. Second, smoothing also increases the spatial covariance of the noise, which can lead on its own to increases in sensitivity of statistical analyses, as it acts to reduce the severity of the required corrections for multiple comparisons [45]. Finally, and perhaps most importantly, smoothing makes the results more robust to residual misalignments between the functional BOLD responses across different subjects. Because most current spatial normalization procedures focus on anatomical homogenization and there is no one-to-one correspondence between anatomy and function, considerable variability in the localization of functional responses across different subjects can be expected to remain. From that perspective, smoothing serves to increase the amount of overlap of functional responses across subjects, increasing the sensitivity of group-level analyses at the cost of reduced spatial specificity [46-48]. Approaches that accommodate natural variability in the spatial localization of function without sacrificing sensitivity or spatial specificity [47,49] offer potential alternatives as increases in the spatial resolution of fMRI data continue to challenge the assumption of a one-to-one mapping between anatomy and function.

## 3. Denoising

After the functional MRI data have been preprocessed, the BOLD signal still contains a considerable amount of variability associated with non-neural sources, such as thermal and drift noise from the scanner, residual effects from subject motion, and respiratory, cardiac, and other physiological effects [50-53]. Properly addressing these noise sources is a crucially important consideration before performing statistical analyses of fMRI data, as without this step the results of these analyses would be highly unreliable, and likely uninterpretable.

The effect of these noise sources on the statistical analysis of fMRI data depends largely on the type of analysis planned. In the analysis of task fMRI data, for example, if the presence of noise covaries with the task (e.g., a subject may move more during a rest condition compared to a demanding task condition) noise can result in increased false positives and reduced replicability, as task-related noise variability is confounded with genuine task-related BOLD responses [54-55]. Even if motion was independent of the task, spurious task correlations can still influence the results, as the magnitude of motion-related variance is often considerably larger than that of task-related variance. Last, even if motion was independent of the task and there were no spurious task-correlations in the sample, the presence of noise would still act to increase the residual BOLD signal variability, reducing the power of the statistical analyses and limiting replicability. Noise can also affect the form of the spatial covariance of the residual BOLD signal, potentially affecting the validity of cluster-level statistics and other procedures that rely on a priori characterizations of this covariance [56]. Last, the presence or relative influence of some of these noise sources may vary from subject to subject (e.g., depending on their relative levels of motion), potentially affecting the validity of common statistical assumptions in group analyses such as homoscedasticity of the residual BOLD signal. In the context of functional connectivity analysis of fMRI data, the effects of noise are even more damaging, as noise sources are often highly correlated across different areas, biasing any measures of BOLD signal correlation between regions [57]. These biases ultimately increase the rate of false positives and challenge the validity of functional connectivity statistical analyses, as a shared noise source between two regions is confounded with genuine functional connectivity between them.

Denoising steps are aimed at directly removing or reducing the influence of these noise sources on the BOLD signal timeseries. They are performed either directly as additional processing steps applied to the functional data before statistical analyses (e.g., removing noise components from the functional data), or jointly as part of an explicit noise control strategy during the statistical analyses (e.g. entering additional noise covariates in a first-level general linear model analysis of task data). Denoising is often performed separately on each individual functional run or session, as it is assumed that the impact of different noise sources on the BOLD signal may vary between runs.

Most denoising steps start by defining one or several time series $g_k[n]$ characterizing some potential noise sources. It is assumed that the BOLD timeseries $f_m[n]$ at each individual voxel $m$ will be contaminated by a linear combination of these noise sources, although individual voxels may naturally differ in how exactly different noise sources affect them. Under these assumptions the influence of each noise source on the BOLD signal can then be estimated and removed in a single step using multivariate linear regression:

$$\boldsymbol{F}^* = \boldsymbol{F} - \boldsymbol{G} \cdot (\boldsymbol{G}^t \cdot \boldsymbol{G})^{-1} \cdot \boldsymbol{G}^t \cdot \boldsymbol{F}$$

where $\boldsymbol{F}$ is a scans-by-voxels matrix with values $f_m[n]$, and $\boldsymbol{G}$ is a scans-by-sources matrix with values $g_k[n]$. This step projects the BOLD timeseries into a subspace orthogonal to the noise components so that statistical analyses can then proceed within this subspace where the presence of noise is considerably reduced or eliminated.

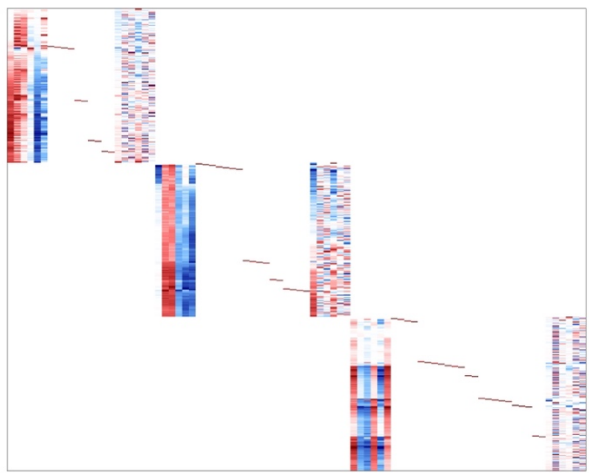

**Fig. 6** Example image of noise matrix $\boldsymbol{G}$ defining noise sources across three different functional runs (block diagonal structure with three blocks, one per run). Each block contains 6 time series associated with subject motion,

a variable number of time series identifying potential outlier scans, and 6 additional time series associated with component-based correction components.

Alternatively, the noise matrix $G$ can also be directly used during the statistical analyses as an explicit control covariate of the BOLD timeseries, e.g., in a first-level general linear model of the form:

$$F = [X\ G] \cdot \begin{bmatrix} \widetilde{B} \\ \widetilde{C} \end{bmatrix} + \varepsilon$$

where the effects $\widetilde{B}$ characterizing the association between a model design matrix $X$ and the functional data $F$ can be estimated while controlling for the influence of known noise components $G$. The two approaches can also be combined, for example by first denoising the functional data using a set of noise components $G$ and then entering the same matrix $G$ as a control covariate in the statistical analysis to allow it to correctly infer the effective degrees of freedom in the denoised data.

The next sections describe some of the most widely used forms of denoising fMRI data, including motion regression, scrubbing, component-based corrections, and filtering. Other popular forms of denoising that require additional measurements, such as RETROICOR which uses respiratory and cardiac signals recorded during the scanner to address physiological noise sources [58], or ME-ICA which requires a multi-echo EPI acquisition and uses the differences in images acquired at different echo times to identify responses of potential neural origin [59-61], are not covered here.

### 3.1 Motion regression

A common way to address residual motion effects in the data is to remove from the BOLD signal time series at each individual voxel any component that may be temporally correlated with common indicators of subject motion. A natural indicator of subject motion affecting an individual voxel in the image could be the changing position of this voxel inside the scanner. This leads to a simple first-order approximation characterizing the range of possible motion effects on the BOLD signal at this voxel as linear combinations of three timeseries representing its changing 3D coordinates.

For simplicity, instead of a separate representation of motion effects at each voxel, a common representation across all voxels may be preferable. A simple common representation is provided by 12 timeseries representing each of the elements of the $M_n$ affine transformation matrices, which, by definition, are linearly related to the 3D coordinates of every individual voxel and hence to the modeled effects of motion on the BOLD signal. An even simpler representation is provided by 6 timeseries representing only three translation and three rotation parameters associated with the same $M_n$ matrices, as only rigid-body transformations are used to define these matrices (see motion correction section above). While not identical, for relatively moderate amounts of subject motion, the 3D coordinates of each voxel can be approximated by linear combinations of these 6 "translation/rotation" timeseries almost as well as by linear combinations of the 12 "affine matrix" timeseries. This leads to the most common form of motion regression of the BOLD signal, consisting of noise components $g_k[n]$ defined using the timeseries associated with each of the six motion parameters estimated during the motion correction preprocessing step: three timeseries indicating subject translations along the x, y, and z directions and three timeseries indicating rotations around the x, y, and z axes. Because the interest is in linear combinations of these timeseries, the scale, order, or units of these timeseries are unimportant. Despite their simplicity, these 6 motion timeseries explain an unexpectedly large proportion of the total variance in the BOLD signal consistently across different voxels and different subjects (see Figure 7).

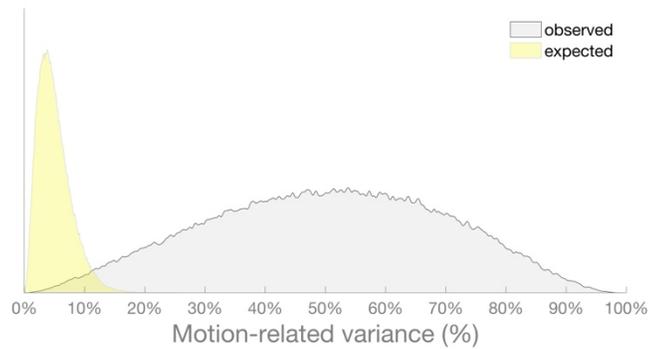

**Fig. 7** Proportion of the total temporal variance in the BOLD signal that is correlated with motion (6 parameters), sampled separately across 1000 random voxels within the gray matter on 198 different subjects. For reference, the proportion of motion-related variance observed in this sample (shaded in gray) is compared to the proportion that could be expected by chance (in yellow).

In addition to these original six timeseries, one or several nonlinear transformations of the same timeseries are also sometimes included, such as first- or higher-order derivatives, one or multiple temporal shifts, and second- or higher-order powers. These additional terms can be used to extend beyond a simple first-order approximation between subject motion and BOLD signal changes, fitting a more complex linear expansion of the potentially nonlinear effects of motion on the BOLD signal [13].

### 3.2 Artifact detection and scrubbing

Individual outlier scans, either as a consequence of extreme sudden subject motion, hardware instabilities, or other abrupt events, can strongly influence the results of statistical analyses of fMRI data.

Motion-related outlier scans can be identified as supra-threshold values in framewise displacement (FD), a composite measure of subject motion computed at each individual scan. Framewise displacement timeseries are computed from the difference in the subject's head position between two consecutive scans, as estimated during the motion correction preprocessing step. There are several similar definitions of framewise displacement, varying in how exactly a summary measure of displacement is computed between two scans. Some of these definitions are simple aggregations of motion parameters, such as C-PAC FD, defined as the sum of the absolute values of the three translation parameters plus the three rotations in radians multiplied by 50 [62], while others have a clearer geometrical interpretation, such as FSL FD, defined as the root mean square of the change in position of all points within an 80 mm sphere undergoing the same rotations and translations as the subject's head [63], or CONN FD, defined as the maximum change in the position of six points placed at the centers of each face in a bounding box encompassing the brain and undergoing the same rotations and translations as the subject's head [57]. Thresholds to identify outlier scans from FD timeseries vary from study to study, typically ranging between 0.5 mm and 2 mm.

In order to capture other events not necessarily related to subject motion, outlier scans can also be identified from measures of scan-to-scan changes in the BOLD signal timeseries. Several similar measures of BOLD signal change have also been proposed and used, such as global signal change (GSC), defined as the absolute difference in global BOLD signal (average signal across all voxels within the brain) between two consecutive scans [57], or DVARS, defined as the root mean square of the difference in BOLD signal between two consecutive scans computed over all voxels within the brain [62]. Thresholds to identify outlier scans from BOLD signal change measures also vary from study to study, with DVARS above 0.5% or GSC above 3 standard deviations being relatively conservative values.

Combining the two approaches above, outlier scans can be identified as scans with suprathreshold values in either FD or BOLD signal change measures. Because outliers are identified from measures of change between two consecutive scans, it is prudent to treat both scans as potential outliers, with the possibility of additionally removing other following images in cases where the effect of these outlier events may be suspected to last multiple scans. In addition to motion and BOLD signal changes, in some cases, it may be useful to identify additional potential outlier scans from different sources of evidence, which may be specific to a study or fMRI acquisition details. For example, to address potential instabilities in the BOLD signal during the first few scans, which can be caused by a transient magnetization state of the tissue at the beginning of a functional acquisition run, it is not uncommon to manually select a fixed number of initial scans as potential outliers so that their effect is removed together with that of other outliers.

Once outlier scans are identified, their effect on the BOLD signal is removed using a technique known as censoring or scrubbing, where one individual noise component $g_k[n]$ is defined for each individual outlier scan, taking a value of 1 for the outlier scan and 0 for all other scans. Using these noise components, the standard linear regression denoising procedure will entirely remove the influence of the identified outlier scans from the BOLD signal timeseries without disrupting the continuity of the data. As with other denoising strategies, scrubbing can be performed before statistical analyses, as part of an explicit denoising step, or integrated with the statistical analyses by using these noise components as control variables.

### 3.3 BOLD noise modeling

Since noise is explicitly defined as variability in the BOLD signal that is associated with non-neural sources, it is natural to look at areas in the brain where there are no neurons and analyze the BOLD signal at those areas in an attempt to find a richer characterization of noise. One such approach is anatomical component-based correction (aCompCor), which performs a principal component analysis of the BOLD signal within white matter and cerebrospinal fluid areas (CSF) to extract a number of components best characterizing the variability in the BOLD signal within these areas [64-66]. White matter and CSF masks are usually heavily eroded to minimize partial volume effects and the potential contamination of the resulting components with signals of neural origin. Principal components estimated from white matter and CSF areas are useful to characterize physiological noise as well as residual motion-related variance in the rest of the brain. For example, in gray matter areas, it is common to find that a very large proportion of the total variance in the BOLD signal is explained by a small set of CompCor noise components estimated from white matter and CSF areas (see Figure 8).

To remove the influence of these noise sources across all voxels, principal components from CompCor can be entered directly as noise components $g_k[n]$ in the standard linear regression denoising procedure. When combining CompCor with other denoising procedures, it is useful to compute principal components that are orthogonal to other already identified noise components, such as those estimated from motion parameters or from potential outlier scans, to focus the resulting CompCor components on additional, still unidentified, sources of noise-related variance.

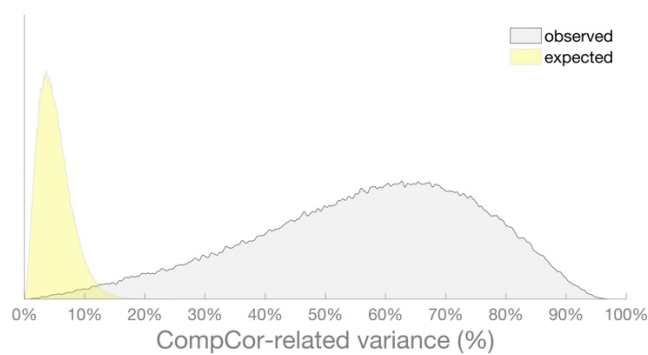

**Fig. 8** Proportion of the total temporal variance in the BOLD signal that is correlated with noise (6 CompCor parameters), sampled across 1000 random voxels within the gray matter on 198 different subjects. For reference, the proportion of noise-related variance observed in this sample (shaded in gray) is compared to the proportion that could be expected by chance (in yellow).

Another similar approach uses independent component analyses (ICA) instead of principal component analysis to extract meaningful components from the BOLD signal. Traditionally, ICA-based denoising uses the BOLD signal from the entire brain, rather than from a limited set of areas such as white matter, to estimate a number of independent components. Because of this, it is necessary to identify among all of the extracted components those that most likely characterize noise sources. This can be done manually, although the process can be time consuming and requires a certain level of expertise, or automatically, using classifiers that are trained to discriminate noise components from other components related to neural sources [67-70]. Features that are frequently found to be useful are based on properties of the time series associated with each potential noise component, such as the relative power in higher frequencies, the presence of spikes, or the correlation with motion or other known noise sources, as well as properties of the spatial distribution of each component, such as the relative overlap with non-gray matter areas.

### 3.4 Frequency filtering

The last step when denoising fMRI data is to filter the BOLD signal time series within a frequency window of interest. The choice of filter depends on the planned statistical analyses. In task-based analyses, where high-frequency content may be essential, a simple high-pass filter with a cutoff frequency of approximately 0.01 Hz can be used, while in functional connectivity analyses, where the interest is often in low-frequency fluctuations, a bandpass filter between 0.01 Hz and 0.10 Hz is often preferred.

Similar to spatial smoothing, frequency filtering serves multiple purposes. It can be used in the context of a classical matched filter approach as a way to remove extremely low frequencies in the BOLD signal, which can be dominated by scanner drift effects or 1/f noise. It can also be used in simultaneous multislice (SMS) acquisitions with fast repetition times (e.g. below 250 ms) to remove high-frequency respiratory and cardiac noise from the BOLD signal. With a sufficiently high sampling rate of fMRI data acquisition, respiratory and cardiac effects on the BOLD signal can often be concentrated and easily identifiable at approximately 0.3 Hz and 1.3 Hz, while with slower acquisition sampling rates, the same effects would appear aliased over the entire sampled frequency range. In other contexts, beyond denoising, filtering can also be used to focus the statistical analyses on a specific frequency window or to analyze the frequency dependency of different measures of interest, such as task responses or functional connectivity.

Similar to other denoising steps, which are typically applied simultaneously in a single linear regression step combining all noise components $g_k[n]$, filtering can also be implemented simultaneously by defining as additional noise components a set of sine and cosine temporal functions spanning all frequencies *outside* of the desired frequency window (an approach referred to as *simult*, simultaneous regression and filtering). This approach is often found to produce similar results to an alternative sequential approach, where bandpass filtering is applied as a separate denoising step after a linear regression step addressing all other noise sources (an approach named RegBP, indicating regression followed by filtering) [71]. However, there are some practical limitations of *simult* that are not shared by RegBP, which may make the latter a preferred approach in many circumstances. First, *simult* can only accommodate filters with flat frequency response, excluding the application of other more general filters, such as Butterworth or similar filters with compact temporal support. Second, *simult* will estimate and remove the effect of other noise components $g_k[n]$ separately within the filter bandpass window, disregarding any information outside of this window, which may be the desired behavior when the effect of these noise components is expected to vary across different frequency windows, but it may also be an unnecessary or wasteful limitation when the effect of these noise components has a narrow temporal support (e.g., the effect of identified outlier scans used in scrubbing).

It is interesting to note that a RegBP approach where noise components $g_k[n]$ are pre-filtered is exactly equivalent to a simultaneous approach. This affords a simple generalization of RegBP encompassing both the traditional RegBP and *simult* approaches where, rather than prefiltering all noise components to implement a standard *simult* approach, or none of them to implement a standard RegBP approach, it is also possible to choose specifically which noise component(s) $g_k[n]$ may be desirable/reasonable to prefilter [57]. This choice, for example, can be made separately for different noise sources based on considerations regarding the nature of the association between each noise source and the BOLD signal. The general recommendation when using a generalized RegBP approach is to filter a particular noise component only if its effect on the BOLD signal can be expected to change depending on the frequency of the noise (particularly when comparing frequencies passed vs. those stopped by the filter).

## 4. Quality Control

After preprocessing and denoising the fMRI data and *before* performing statistical analyses, it is crucial to perform thorough quality control (QC) procedures aimed at identifying potential problems in the data or in any of the preprocessing and denoising steps [72-75]. Failure to do this will inevitably lead to problems later on, either in the form of results that do not replicate, results that are strongly affected by the presence of outliers, or results that are artificially matched to our expectations (e.g. by only addressing problems in the data after failing to observe an expected result).

The next section describes some quality control measures that can be computed automatically and that can help flag potential problems in the data. While it may be tempting to purely rely on such measures for quality control, visual inspection of raw functional and structural images as well as representative images after each of the preprocessing steps is an equally important part of any proper quality control pipeline. Obtaining fMRI data is time consuming, expensive, and demanding for our subjects. Visual inspection is but a small effort that can not only help identify problems in the data but can also help researchers familiarize themselves with the specificities of their data.

Among standard visual inspection procedures, displaying raw anatomical and functional images can help detect potential acquisition artifacts, such as ghosting, Gibbs artifacts, or spatial distortions caused by implants or other objects. It can also help evaluate the nature and extent of magnetic susceptibility distortions in the data or detect possible encoding or data conversion problems. After motion correction, displaying the first and last scan in each functional run from a subject can help detect potential problems in realignment or inter-run coregistration. After anatomical-functional coregistration, superimposing or quickly alternating between anatomical and functional slices while visually evaluating the match between anatomical features in the two images can help assess the accuracy of the intermodality coregistration, which can in turn help guide other choices of preprocessing steps. Similarly, after intersubject normalization, displaying the functional or anatomical images superimposed with an a priori gray matter tissue probability map can help detect potential problems in the normalization procedure, or evaluate the appropriateness of the chosen templates for the target population. Across multiple subjects, displaying a mosaic or alternating view of an individual functional or anatomical slice spanning all subjects can also help quickly identify potential outlier subjects. Last, displaying descriptive measures related to the specific statistical analyses planned, such as task-related response distributions for each subject in task-based analyses, analysis masks in voxel-level group analyses [76], or voxel-to-voxel correlation distributions for each subject in functional connectivity analyses [57], can also help identify potential otherwise unspecific problems in the data.

### 4.1 Subject-level quality control measures

There is a variety of quality control measures evaluating different aspects of the fMRI data at the level of individual subjects or individual runs. Their main application is to help identify problems in the data, possibly indicating potential outlier subjects or runs within a study. In that context, it is their relative values compared to other runs or subjects within the same study what carries the most useful information. While it may be tempting to use the absolute values of these measures to establish universal thresholds of data quality across studies, or to compare the quality of different datasets or different preprocessing procedures, this should always be done with care as many of these measures can be expected to show variability with differences in acquisition parameters or differences in the details of the analysis pipeline which may not afford such a simple interpretation as when those aspects are common across all subjects or runs. Descriptive analyses of these QC measures can be used to quickly identify potential outlier subjects or runs, for example (see Figure 9) by identifying those values above the upper quartile plus 1.5 times the interquartile range for mild outliers, or the upper quartile plus 3 times the interquartile range for extreme outliers (or the equivalent lower quartile thresholds when lower values in a QC measure are indicative of potential problems).

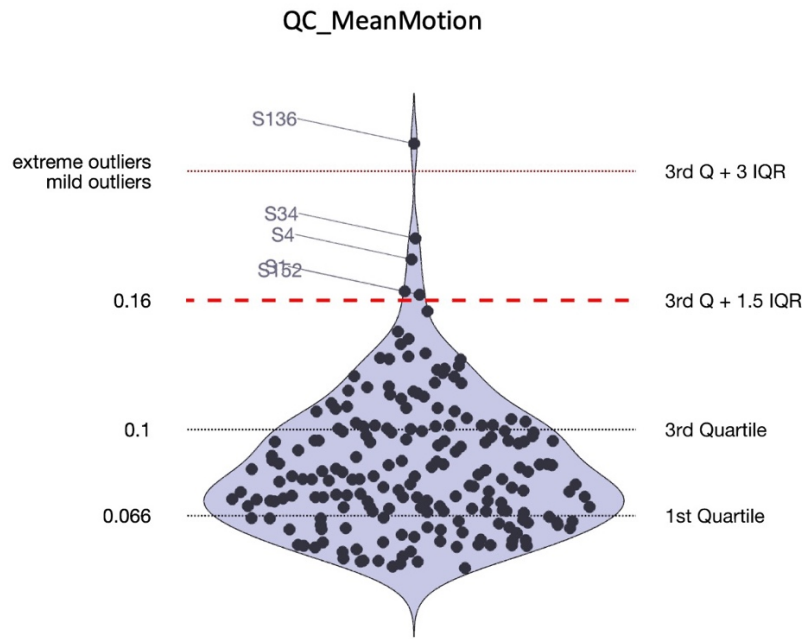

**Fig. 9** Example of the distribution of subject motion values (average FD) across all subjects in a study.

**Subject motion**: QC measures quantifying the amount or severity of motion for each subject can be computed by averaging framewise displacement (FD) across all scans for each subject. If outlier scans have been identified, a more sensitive measure can be computed by averaging only across valid scans (disregarding outlier scans) and reporting the number of outlier scans separately [6].

**BOLD signal stability**: QC measures quantifying BOLD signal stability can be computed by averaging BOLD signal change measures, such as DVARS or GSC. Similar to subject motion measures, computing this average only among valid scans leads to a measure more representative of the quality of the data after scrubbing (with higher values indicating potential problems). A similar measure (BOLD std) directly computes the standard deviation of the BOLD signal at each voxel and then aggregates across all voxels within the brain. Another related measure is global correlation (GCOR), which first aggregates across all voxels their standardized unit-variance BOLD signals and then computes the standard deviation of the resulting global signal [77]. Yet another similar measure (but with inverse interpretation) is temporal signal-to-noise ratio (tSNR) [78], which represents the ratio between the average value of the BOLD signal and its standard deviation.

**Outlier scans**: the total number of scans identified as potential outliers by the artifact detection procedure is also a useful QC measure quantifying the overall quality of the functional data. Similarly, proportion of either valid or outlier scans can also be used to equalize these measures across subjects when the number and length of functional acquisitions differ among them.

**Effective degrees of freedom**: measures that attempt to quantify the effective temporal degrees of freedom [79] of the functional time series after denoising are useful to identify potential heteroscedasticity across subjects as well as potential overdenoising scenarios for individual subjects (where insufficient residual time series variability remains after removal of all of the identified noise components). Approximate measures such as those computed as the product of the number of scans minus the number of noise components, multiplied by the proportion of frequencies kept by the frequency filter, are also typically sufficient for QC purposes.

**Anatomical-functional overlap**: QC measures that quantify the accuracy of coregistration between the functional and anatomical data can be computed from measures of match or overlap between the two images, for example mutual information between the two images [19], or dice coefficients between gray matter masks computed separately from each image modality [6].

**Normalization accuracy**: similar to functional-anatomical overlap measures, QC measures that quantify the accuracy of spatial normalization of functional and anatomical images can be computed from measures of overlap between these images and their respective template images, for example dice coefficients between gray matter masks computed from each image modality and a template gray matter probability map (see Figure 10) [6].

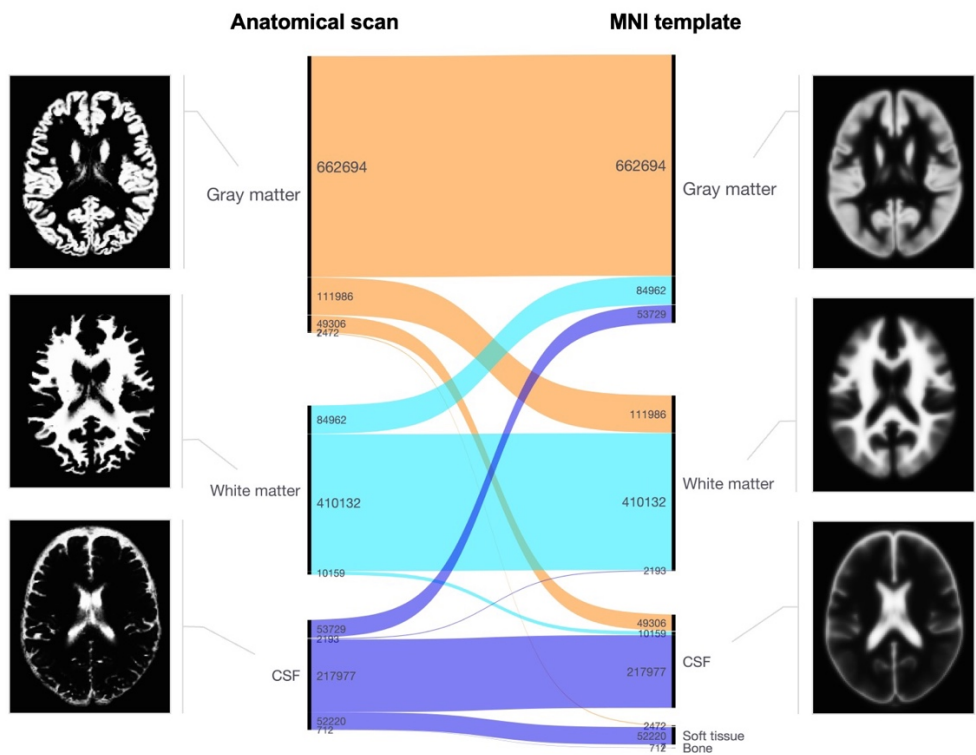

**Fig. 10** Example of overlap between tissue masks in a subject's anatomical scan after normalization (left), and a reference tissue probability mask in MNI-space (right).

### 4.2 Group-level quality control measures

While not a substitute for detailed subject-level quality control procedures, procedures that evaluate sources of intersubject variability in an entire dataset to provide overall quality measures can also be useful to help researchers identify and resolve potential problems in the data before proceeding to any planned statistical analyses. Group-level quality control measures are also more likely to generalize across different studies and acquisition details, so they can be useful when comparing different datasets or different pipelines.

A recommended group-level QC practice is to evaluate the correlation across subjects between subject-level quality control measures and functional measures of interest. The choice of functional measures of interest is typically matched to those considered in the planned statistical analysis. For example in functional activation analyses, group-level tests evaluating the association across subjects between subject-motion severity and task-related responses can be helpful to identify the presence of potential confounder effects.

Because many noise sources affect similarly large portions of the brain, their presence tends to affect the spatial correlation structure of the BOLD signal in a particularly consistent manner. Accordingly, one of the most promising group-level QC approaches, QC-FC correlations [80] borrowed from the functional connectivity literature, evaluates changes in the spatial correlation structure of the BOLD data that may covary with subject-level quality control measures. In particular, bivariate temporal correlations between the BOLD signals from different regions are used to sample the spatial correlation structure of the BOLD data. These values are computed from a large sample of different connections (e.g., all possible pairs among 1000 voxels or small ROIs distributed across the entire brain) to provide a relatively dense sample of the entire voxel-to-voxel correlation structure. The QC-FC intersubject correlation is then computed separately for each of these connections. Departures in the shape of the observed distribution of QC-FC intersubject correlation values from that expected by chance are quantified using a Kolmogorov-Smirnov-like measure of distributional distance, and match levels above 95% are considered as indicative of negligible modulations in the BOLD signal correlation structure with subject motion [57]. The same analyses can be repeated for different QC measures to rule out other potential sources of noise. Common QC measures used in QC-FC analyses include average framewise displacement, number of potential outlier scans found during the artifact detection step, and the proportion of valid scans for each subject. In contrast, other QC measures more directly related to BOLD signal properties, such as BOLD signal stability, are not considered as useful, as non-spurious associations between these measures and the observed BOLD spatial correlation structure may naturally exist in otherwise uncontaminated data.

While QC-FC analyses were originally proposed as a group-level quality control measure in the context of functional connectivity analyses, their applicability extends to any other statistical analyses of fMRI data, and they can be generally used to identify residual problems in the BOLD data regardless of the planned statistical analyses.

### 5. Summary and Conclusions

This chapter described preprocessing, denoising, and quality assurance steps involved in preparing fMRI data for statistical analyses.

Preprocessing steps focus on spatial properties of both the functional and anatomical images. They include susceptibility distortion correction, where inhomogeneities in the magnetic field created by the presence of a subject ultimately produce characteristic spatial distortions in the phase encoded direction of the EPI data, intermodality coregistration, where functional, anatomical, or other imaging modalities are aligned in a common reference frame, intersubject normalization, addressing anatomical differences between subjects, and smoothing, aimed at increasing SNR and compensating for residual intersubject heterogeneity in the localization of functional activation.

Denoising steps focus on identifying and removing remaining sources of temporal variability in the BOLD signal. These steps include the regression of identified factors characterizing noise in the BOLD signal, such as motion regression, scrubbing, component-based corrections, and frequency filtering. The identified factors still account for the vast majority of the observed variability in the BOLD signal after preprocessing the data, so a careful control of these factors during denoising can help substantially improve power and replicability of any subsequent statistical analyses.

Last, quality control procedures are an essential part of preparing fMRI data for statistical analyses. This chapter described several procedures focusing on visual inspection of the functional and anatomical images, as well as several QC measures that can be used to evaluate some of the aspects affecting the quality of each individual subject's data, such as average framewise displacement, average global signal change, or the proportion of artifactual scans. The section ends describing QC-FC, a promising group-level QC procedure that can be used to obtain a single omnibus measure evaluating the presence of residual motion effects or other harmful factors on the data before proceeding to statistical analyses.

The importance of thorough preprocessing, denoising, and quality control procedures before statistical analyses of the fMRI data cannot be overstated. The common adage "garbage in, garbage out" applies, as no matter how sophisticated our statistical analyses, they will produce only noise if the data is not properly curated. More alarmingly, noise output by statistical analyses can often look like genuine results, not only through the well understood loop of publication biases and a field overreliance on statistical significance, but also simply because the causes of noise are often proxies to other meaningful constructs (e.g. subject motion as a proxy for age or for disease severity, cardiac or respiratory rate for anxiety, etc.), resulting in low replicability and a high risk of misinterpreting the results of fMRI analyses. While there is no guaranteed approach or pipeline that will result in "clean" data, the preprocessing, denoising, and quality assurance steps described in this chapter can help maximize the chances that subsequent analyses of fMRI data will produce meaningful and replicable results.

**References**

[1] Poynton, C., Jenkinson, M., Pierpaoli, C., & III, W. W. (2009). Fieldmap-Free Retrospective Registration and Distortion Correction for EPI-based Diffusion-Weighted Imaging. In Proc. Intl. Soc. Mag. Reson. Med (Vol. 17).

[2] Jezzard P, Clare S (1999) Sources of distor- tion in functional MRI data. Hum Brain Mapp 8(2):80–85

[3] Hutton C, Bork A, Josephs O, Deichmann R, Ashburner J, Turner R (2002) Image distor- tion correction in fMRI: a quantitative evalua- tion. Neuroimage 16(1):217–240

[4] Jenkinson, M., Bannister, P., Brady, J. M. and Smith, S. M. Improved Optimisation for the Robust and Accurate Linear Registration and Motion Correction of Brain Images. NeuroImage, 17(2), 825-841, 2002.

[5] Penny, W. D., Friston, K. J., Ashburner, J. T., Kiebel, S. J., & Nichols, T. E. (Eds.). (2011). Statistical parametric mapping: the analysis of functional brain images. Elsevier.

[6] Nieto-Castanon, A. (2021). CONN functional connectivity toolbox (RRID:SCR_009550), Version 21. Hilbert Press. doi:10.56441/hilbertpress.2161.7292

[7] Jenkinson, M., Beckmann, C. F., Behrens, T. E., Woolrich, M. W., & Smith, S. M. (2012). Fsl. Neuroimage, 62(2), 782-790.

[8] Cox, R. W., & Hyde, J. S. (1997). Software tools for analysis and visualization of fMRI data. NMR in Biomedicine: An International Journal Devoted to the Development and Application of Magnetic Resonance In Vivo, 10(4-5), 171-178.

[9] Esteban, O., Markiewicz, C. J., Blair, R. W., Moodie, C. A., Isik, A. I., Erramuzpe, A., ... & Gorgolewski, K. J. (2019). fMRIPrep: a robust preprocessing pipeline for functional MRI. Nature methods, 16(1), 111-116.

[10] Andersson, J. L., Hutton, C., Ashburner, J., Turner, R., & Friston, K. (2001). Modeling geometric deformations in EPI time series. Neuroimage, 13(5), 903-919.

[11] Bannister, P. R., Brady, J. M., & Jenkinson, M. (2004). TIGER–a new model for spatio-temporal realignment of fMRI data. In Computer Vision and Mathematical Methods in Medical and Biomedical Image Analysis (pp. 292-303). Springer, Berlin, Heidelberg.

[12] Beall, E. B., & Lowe, M. J. (2014). SimPACE: generating simulated motion corrupted BOLD data with synthetic-navigated acquisition for the development and evaluation of SLOMOCO: a new, highly effective slicewise motion correction. Neuroimage, 101, 21-34.

[13] Friston KJ, Williams S, Howard R, Frackowiak RSJ, Turner R (1996) Movement-related effects in fMRI time-series. Magn Reson Med 35:346–355

[14] Beall, E.B., Lowe, M.J. (2015). Retrospective nonlinear spin history motion artifact modeling and correction with SLOMOCO. Proc. Int. Soc. Magn. Reson. Med. 23, 2104.

[15] Thévenaz P, Blu T, Unser M (2000) Interpolation revisited. IEEE Trans Med Imaging 19(7):739–758

[16] Henson, R. N. A., Buechel, C., Josephs, O., & Friston, K. J. (1999). The slice-timing problem in event-related fMRI. NeuroImage, 9, 125

[17] Friston, K. J., Fletcher, P., Josephs, O., Holmes, A. N. D. R. E. W., Rugg, M. D., & Turner, R. (1998). Event-related fMRI: characterizing differential responses. Neuroimage, 7(1), 30-40.

[18] Parker, D. B., & Razlighi, Q. R. (2019). The benefit of slice timing correction in common fMRI preprocessing pipelines. Frontiers in neuroscience, 13, 821.

[19] Collignon A, Maes F, Delaere D, Vandermeulen D, Suetens P, Marchal G (1995) Automated multi-modality image registration based on information theory. In: Bizais Y, Barillot C, Di Paola R (eds) Proc Information Processing in Medical Imaging (IPMI). Kluwer Academic, Dordrecht, pp 263–274

[20] Wells WM III, Viola P, Atsumi H, Nakajima S, Kikinis R (1996) Multi-modal volume registra- tion by maximisation of mutual information. Med Image Anal 1(1):35–51

[21] Studholme, Hill & Hawkes (1998). A normalized entropy measure of 3-D medical image alignment. Proc. Medical Imaging 1998, vol. 3338, San Diego, CA, pp. 132-143

[22] Jenkinson, M. and Smith, S. M. (2001). A Global Optimisation Method for Robust Affine Registration of Brain Images. Medical Image Analysis, 5(2), 143-156.

[23] Gonzalez-Castillo, J., Duthie, K. N., Saad, Z. S., Chu, C., Bandettini, P. A., & Luh, W. M. (2013). Effects of image contrast on functional MRI image registration. NeuroImage, 67, 163-174.

[24] Fedorenko, E., Hsieh, P. J., Nieto-Castañón, A., Whitfield-Gabrieli, S., & Kanwisher, N. (2010). New method for fMRI investigations of language: defining ROIs functionally in individual subjects. Journal of neurophysiology, 104(2), 1177-1194.

[25] Fischl, B., Sereno, M. I., & Dale, A. M. (1999). Cortical surface-based analysis: II: inflation, flattening, and a surface-based coordinate system. Neuroimage, 9(2), 195-207.

[26] Fischl, B., Sereno, M. I., Tootell, R. B., & Dale, A. M. (1999). High-resolution intersubject averaging and a coordinate system for the cortical surface. Human brain mapping, 8(4), 272-284.

[27] Klein, A., Andersson, J., Ardekani, B. A., Ashburner, J., Avants, B., Chiang, M. C., ... & Parsey, R. V. (2009). Evaluation of 14 nonlinear deformation algorithms applied to human brain MRI registration. Neuroimage, 46(3), 786-802.

[28] Avants, B. B., Tustison, N. J., Song, G., Cook, P. A., Klein, A., & Gee, J. C. (2011). A reproducible evaluation of ANTs similarity metric performance in brain image registration. Neuroimage, 54(3), 2033-2044.

[29] Sotiras, A., Davatzikos, C., & Paragios, N. (2013). Deformable medical image registration: A survey. IEEE transactions on medical imaging, 32(7), 1153-1190.

[30] Mazziotta, J., Toga, A., Evans, A., Fox, P., Lancaster, J., Zilles, K., ... & Mazoyer, B. (2001). A probabilistic atlas and reference system for the human brain: International Consortium for Brain Mapping (ICBM). Philosophical Transactions of the Royal Society of London. Series B: Biological Sciences, 356(1412), 1293-1322.

[31] Joshi S, Davis B, Jomier M, Gerig G (2004) Unbiased diffeomorphic atlas construction for computational anatomy. NeuroImage 23:S151–S160

[32] Ashburner, J. and Friston, K. (1997). Multimodal image coregistration and partitioning—a unified framework. Neuroimage, 6(3), 209-217.

[33] Ashburner, J., & Friston, K. J. (2005). Unified segmentation. Neuroimage, 26(3), 839-851.

[34] Bookstein, F. L. (1989). Principal warps: Thin-plate splines and the decomposition of deformations. IEEE Transactions on pattern analysis and machine intelligence, 11(6), 567-585.

[35] Bookstein, F. L. (1997, June). Quadratic variation of deformations. In Biennial International Conference on Information Processing in Medical Imaging (pp. 15-28). Springer, Berlin, Heidelberg.

[36] Woods, R. P., Grafton, S. T., Holmes, C. J., Cherry, S. R., & Mazziotta, J. C. (1998). Automated image registration: I. General methods and intrasubject, intramodality validation. Journal of computer assisted tomography, 22(1), 139-152.

[37] Ashburner, J., & Friston, K. J. (1999). Nonlinear spatial normalization using basis functions. Human brain mapping, 7(4), 254-266.

[38] Rueckert, D., Sonoda, L. I., Hayes, C., Hill, D. L., Leach, M. O., & Hawkes, D. J. (1999). Nonrigid registration using free-form deformations: application to breast MR images. IEEE transactions on medical imaging, 18(8), 712-721.

[39] Thévenaz, P., & Unser, M. (2000). Optimization of mutual information for multiresolution image registration. IEEE transactions on image processing, 9(12), 2083-2099.

[40] Avants, B. B., Tustison, N., & Song, G. (2009). Advanced normalization tools (ANTS). Insight j, 2(365), 1-35.

[41] Ashburner, J. (2007). A fast diffeomorphic image registration algorithm. Neuroimage, 38(1), 95-113.

[42] Ashburner J, Friston KJ (2011) Diffeomorphic registration using geodesic shooting and Gauss–Newton optimisation. NeuroImage 55(3):954–967

[43] Christensen GE, Rabbitt RD, Miller MI, Joshi SC, Grenander U, Coogan TA, Van Essen DC (1995) Topological properties of smooth anatomic maps. In: Bizais Y, Barillot C, Di Paola R (eds) Proc Information Processing in Medical Imaging (IPMI). Kluwer Academic, Dordrecht, pp 101–112

[44] Calhoun, V. D., Wager, T. D., Krishnan, A., Rosch, K. S., Seymour, K. E., Nebel, M. B., ... & Kiehl, K. (2017). The impact of T1 versus EPI spatial normalization templates for fMRI data analyses. Human brain mapping, 38(11), 5331-5342.

[45] Worsley, K. J. (2005). Spatial smoothing of autocorrelations to control the degrees of freedom in fMRI analysis. NeuroImage, 26(2), 635-641.

[46] Mikl, M., Mareček, R., Hluštík, P., Pavlicová, M., Drastich, A., Chlebus, P., ... & Krupa, P. (2008). Effects of spatial smoothing on fMRI group inferences. Magnetic resonance imaging, 26(4), 490-503.

[47] Nieto-Castañón, A., & Fedorenko, E. (2012). Subject-specific functional localizers increase sensitivity and functional resolution of multi-subject analyses. Neuroimage, 63(3), 1646-1669.

[48] Chen, Z., & Calhoun, V. (2018). Effect of spatial smoothing on task fMRI ICA and functional connectivity. Frontiers in neuroscience, 12, 15.

[49] Lindquist, M. A., Loh, J. M., & Yue, Y. R. (2010). Adaptive spatial smoothing of fMRI images. Statistics and its Interface, 3(1), 3-13.

[50] Friston, K. J., Williams, S., Howard, R., Frackowiak, R. S., & Turner, R. (1996). Movement-related effects in fMRI time-series. Magnetic resonance in medicine, 35(3), 346-355.

[51] Bianciardi, M., Fukunaga, M., van Gelderen, P., Horovitz, S. G., de Zwart, J. A., Shmueli, K., & Duyn, J. H. (2009). Sources of functional magnetic resonance imaging signal fluctuations in the human brain at rest: a 7 T study. Magnetic resonance imaging, 27(8), 1019-1029.

[52] Liu, T. T. (2016). Noise contributions to the fMRI signal: An overview. NeuroImage, 143, 141-151.

[53] Caballero-Gaudes, C., & Reynolds, R. C. (2017). Methods for cleaning the BOLD fMRI signal. Neuroimage, 154, 128-149.

[54] Hajnal, J. V., Myers, R., Oatridge, A., Schwieso, J. E., Young, I. R., & Bydder, G. M. (1994). Artifacts due to stimulus correlated motion in functional imaging of the brain. Magnetic resonance in medicine, 31(3), 283-291.

[55] Birn, R. M., Murphy, K., Handwerker, D. A., & Bandettini, P. A. (2009). fMRI in the presence of task-correlated breathing variations. Neuroimage, 47(3), 1092-1104.

[56] Eklund, A., Nichols, T. E., & Knutsson, H. (2016). Cluster failure: Why fMRI inferences for spatial extent have inflated false-positive rates. Proceedings of the national academy of sciences, 113(28), 7900-7905.

[57] Nieto-Castanon, A. (2020). Handbook of functional connectivity Magnetic Resonance Imaging methods in CONN. Hilbert Press.

[58] Glover, G. H., Li, T. Q., & Ress, D. (2000). Image-based method for retrospective correction of physiological motion effects in fMRI: RETROICOR. Magnetic Resonance in Medicine: An Official Journal of the International Society for Magnetic Resonance in Medicine, 44(1), 162-167.

[59] Poser, B. A., Versluis, M. J., Hoogduin, J. M., & Norris, D. G. (2006). BOLD contrast sensitivity enhancement and artifact reduction with multiecho EPI: parallel-acquired inhomogeneity-desensitized fMRI. Magnetic Resonance in Medicine: An Official Journal of the International Society for Magnetic Resonance in Medicine, 55(6), 1227-1235.

[60] Kundu, P., Inati, S. J., Evans, J. W., Luh, W. M., & Bandettini, P. A. (2012). Differentiating BOLD and non-BOLD signals in fMRI time series using multi-echo EPI. Neuroimage, 60(3), 1759-1770.

[61] Kundu, P., Voon, V., Balchandani, P., Lombardo, M. V., Poser, B. A., & Bandettini, P. A. (2017). Multi-echo fMRI: a review of applications in fMRI denoising and analysis of BOLD signals. Neuroimage, 154, 59-80.

[62] Power, J. D., Barnes, K. A., Snyder, A. Z., Schlaggar, B. L., & Petersen, S. E. (2012). Spurious but systematic correlations in functional connectivity MRI networks arise from subject motion. Neuroimage, 59(3), 2142-2154.

[63] Jenkinson, M. (1999). Measuring transformation error by RMS deviation. Studholme, C., Hill, DLG, Hawkes, DJ.

[64] Behzadi, Y., Restom, K., Liau, J., & Liu, T. T. (2007). A component based noise correction method (CompCor) for BOLD and perfusion based fMRI. Neuroimage, 37(1), 90-101.

[65] Chai, X. J., Nieto-Castañón, A., Öngür, D., & Whitfield-Gabrieli, S. (2012). Anticorrelations in resting state networks without global signal regression. Neuroimage, 59(2), 1420-1428.

[66] Jo, H. J., Gotts, S. J., Reynolds, R. C., Bandettini, P. A., Martin, A., Cox, R. W., & Saad, Z. S. (2013). Effective preprocessing procedures virtually eliminate distance-dependent motion artifacts in resting state FMRI. Journal of applied mathematics, 2013.

[67] De Martino, F., Gentile, F., Esposito, F., Balsi, M., Di Salle, F., Goebel, R., & Formisano, E. (2007). Classification of fMRI independent components using IC-fingerprints and support vector machine classifiers. Neuroimage, 34(1), 177-194.

[68] Beall, E. B., & Lowe, M. J. (2007). Isolating physiologic noise sources with independently determined spatial measures. Neuroimage, 37(4), 1286-1300.

[69] Tohka, J., Foerde, K., Aron, A. R., Tom, S. M., Toga, A. W., & Poldrack, R. A. (2008). Automatic independent component labeling for artifact removal in fMRI. Neuroimage, 39(3), 1227-1245.

[70] Beckmann, C. F. (2012). Modelling with independent components. Neuroimage, 62(2), 891-901.

[71] Hallquist, M. N., Hwang, K., & Luna, B. (2013). The nuisance of nuisance regression: spectral misspecification in a common approach to resting-state fMRI preprocessing reintroduces noise and obscures functional connectivity. Neuroimage, 82, 208-225.

[72] Friedman, L., & Glover, G. H. (2006). Report on a multicenter fMRI quality assurance protocol. Journal of Magnetic Resonance Imaging: An Official Journal of the International Society for Magnetic Resonance in Medicine, 23(6), 827-839.

[73] Bennett, C. M., & Miller, M. B. (2010). How reliable are the results from functional magnetic resonance imaging?. Annals of the New York Academy of Sciences, 1191(1), 133-155.

[74] Marcus, D. S., Harms, M. P., Snyder, A. Z., Jenkinson, M., Wilson, J. A., Glasser, M. F., ... & WU-Minn HCP Consortium. (2013). Human Connectome Project informatics: quality control, database services, and data visualization. Neuroimage, 80, 202-219.

[75] Alfaro-Almagro, F., Jenkinson, M., Bangerter, N. K., Andersson, J. L., Griffanti, L., Douaud, G., ... & Smith, S. M. (2018). Image processing and Quality Control for the first 10,000 brain imaging datasets from UK Biobank. Neuroimage, 166, 400-424.

[76] Poldrack, R. A., Fletcher, P. C., Henson, R. N., Worsley, K. J., Brett, M., & Nichols, T. E. (2008). Guidelines for reporting an fMRI study. Neuroimage, 40(2), 409-414.

[77] Saad, Z. S., Reynolds, R. C., Jo, H. J., Gotts, S. J., Chen, G., Martin, A., & Cox, R. W. (2013). Correcting brain-wide correlation differences in resting-state FMRI. Brain connectivity, 3(4), 339-352.

[78] Murphy, K., Bodurka, J., & Bandettini, P. A. (2007). How long to scan? The relationship between fMRI temporal signal to noise ratio and necessary scan duration. Neuroimage, 34(2), 565-574.

[79] Worsley, K. J., & Friston, K. J. (1995). Analysis of fMRI time-series revisited—again. Neuroimage, 2(3), 173-181.

[80] Ciric, R., Wolf, D. H., Power, J. D., Roalf, D. R., Baum, G. L., Ruparel, K., ... & Satterthwaite, T. D. (2017). Benchmarking of participant-level confound regression strategies for the control of motion artifact in studies of functional connectivity. Neuroimage, 154, 174-187.